\documentstyle[11pt,newpasp,twoside,epsf]{article}
\def\edcomment#1{\iffalse\marginpar{\raggedright\sl#1\/}\else\relax\fi}
\reversemarginpar

\begin{document}
\title{Cosmological Simulations with Adaptive Smoothed Particle Hydrodynamics}
\author{Hugo Martel}
\affil{Department of Astronomy, University of Texas, Austin, TX 78712, USA}
\author{Paul R. Shapiro}
\affil{Department of Astronomy, University of Texas, Austin, TX 78712, USA}

\begin{abstract}
We summarize the ideas that led to the
Adaptive Smoothed Particle Hydrodynamics (ASPH) algorithm,
with anisotropic smoothing and shock-tracking.
We then identify a serious new problem for SPH
simulations with shocks and radiative cooling
--- {\it false cooling} --- and discuss a possible solution
based on the shock-tracking ability of ASPH.
\end{abstract}

\section{Introduction}

SPH is the most widely used numerical
method for cosmological simulations with gas dynamics. With our collaborators,
we have developed a new version of SPH, called Adaptive SPH (ASPH),
which addresses some specific limitations of standard SPH
(Shapiro et al. 1996; Owen et al. 1998; Martel \& Shapiro 2002).
For a given number of particles, ASPH resolves much better than
standard SPH whenever anisotropic collapse
or expansion occurs. ASPH also has a shock-tracking algorithm
that can be used to restrict spurious heating by artificial viscosity.
ASPH simulations in 3D of explosions during galaxy formation are
described elsewhere (Martel \& Shapiro 2000, 2001a, b).

\section{Standard SPH vs. Adaptive SPH (ASPH)}

\begin{figure}
\plotone{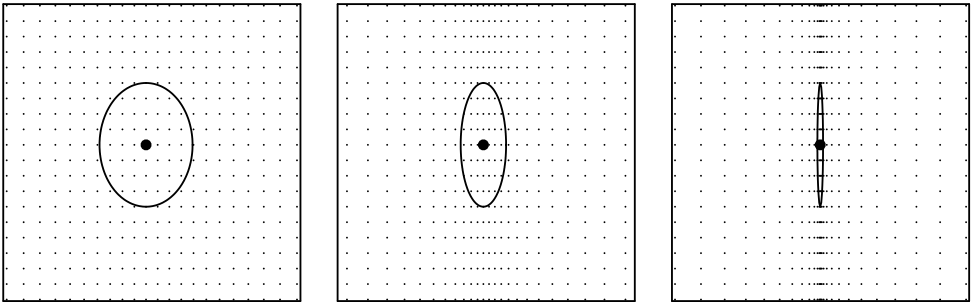}
\caption{Contraction of 2D particle distributions, illustrating the
evolution of the H-tensor. We focus on one particle, represented
by a large dot. The solid curve represents the zone of influence.
Top row: Isotropic contraction with isotropic smoothing.
Middle row: Anisotropic contraction with isotropic smoothing.
Bottom row: Anisotropic contraction with anisotropic smoothing.
}
\end{figure}

For kernel smoothing in standard SPH 
to be accurate, the smoothing length $h$ must
be a few times the mean particle spacing $\Delta x$. If
$h\gg\Delta x$, spatial variations in the fluid quantities
are oversmoothed, leading to a loss of resolution; if
$h<\Delta x$, particles lose contact with their nearest neighbors, 
resulting in a loss of accuracy. Since the mean particle spacing
varies with time and space, each particle must carry its own smoothing
length, which varies with time to reflect the expansion or contraction
of the fluid. This is illustrated in Figure~1. The top row
shows a 2D distribution of particles contracting isotropically.
We focus on one particle, shown by a large dot. The smoothing length
$h$ of that particle defines a {\it zone of influence}, indicated by
a circle. As the fluid contracts, $h$ is reduced in proportion to
the mean particle spacing.
The second row illustrates the case of an anisotropic contraction,
the
planar collapse of a sinusoidal density perturbation, leading to
the formation of a caustic.
In the direction of collapse, the
smoothing length does not shrink fast enough to keep up with the contraction
of the fluid,
and eventually greatly exceeds the mean particle spacing,
leading to poor resolution. In the transverse direction,
the smoothing length shrinks but the fluid does not contract, and horizontal
rows of particles progressively lose contact with one another.
The third row illustrates the ASPH approach. The smoothing is
anisotropic. The smoothing length is replaced by an {\it H-tensor}
which defines an elliptical zone of influence (ellipsoidal in 3D). This
zone of influence 
deforms and rotates to follow the deformation of the fluid.
The smoothing length is now direction dependent, and remains proportional
to the mean particle spacing {\it in any direction}.

\begin{figure}
\plotone{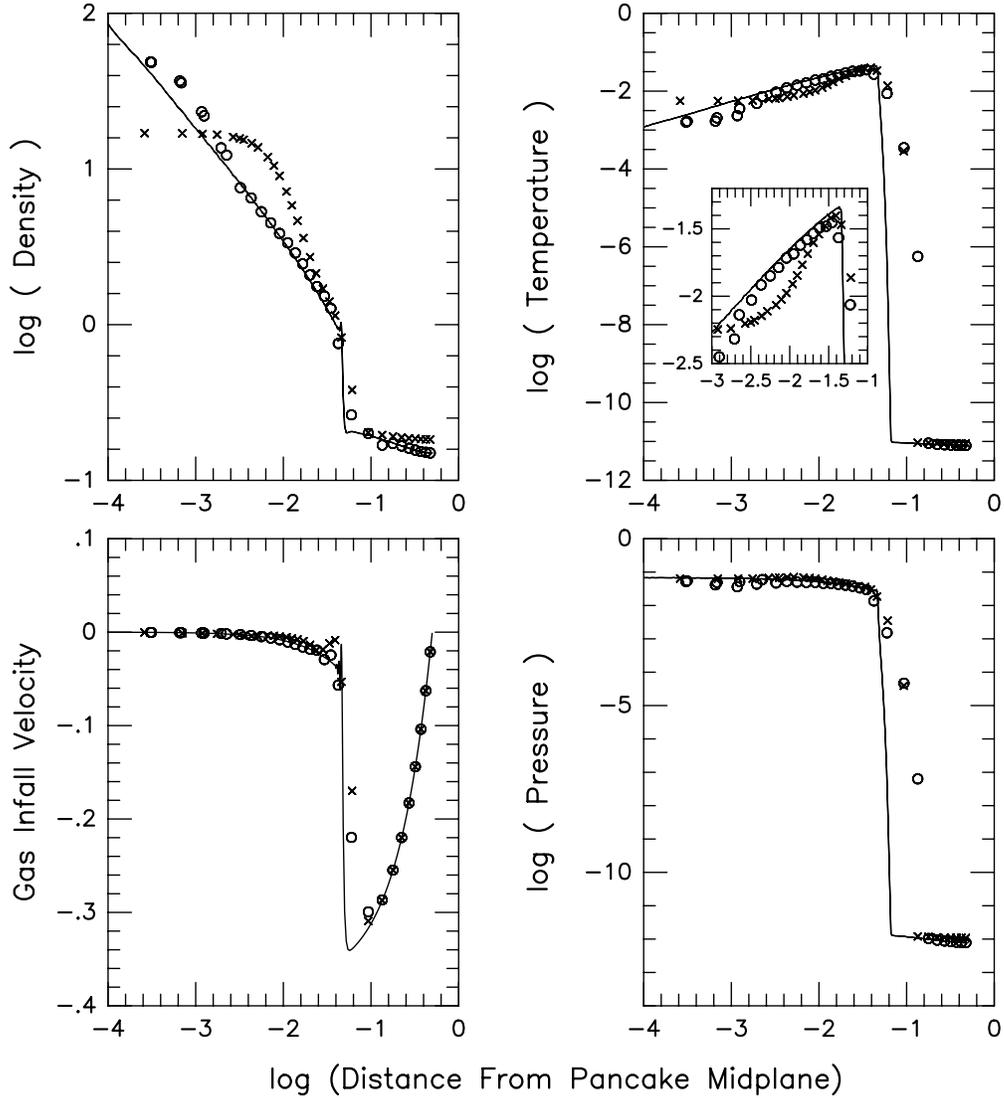}
\caption{Pancake collapse at $a/a_c=2.333$. 
Top left panel: gas density profile.
Top right panel: temperature profile.
Bottom left panel: velocity profile.
Bottom right
panel: pressure profile. Solid curves, crosses, and
circles show the exact, standard SPH, and ASPH 
results, respectively, with distances in units
of $\lambda_p$, the pancake wavelength.
The inset in the top right panel shows an enlargement
of the immediate postshock region.
}
\end{figure}

Artificial
viscosity is necessary to allow the formation of shocks, but 
can lead to spurious preheating of
gas contracting supersonically far from any shock. In ASPH,
the evolution of the H-tensors can be used to track the location of 
shocks and apply artificial viscosity selectively.
This is illustrated in the third row of Figure~1.
In ASPH, viscous heating is initially turned off.
As the system evolves, the smoothing ellipsoid
flattens, with one axis approaching zero
length. ASPH uses this to determine when a particle is about to be
shocked and turns viscous heating on for that particle. 
This restricts viscous heating to particles encountering shocks, as
needed.

To illustrate how ASPH achieves higher
resolution than standard SPH, we focus on
a stringent test, the planar collapse of a sinusoidal plane-wave
density perturbation, the cosmological pancake problem.
The fluctuation grows 
from linear to nonlinear amplitude and forms a caustic in the dark matter
distribution in the plane of symmetry at scale factor $a=a_c$,
with strong accretion
shocks located on each side of the central plane. We
use equal-mass dark matter and gas particles, in a universe
with $\Omega=1$ and $\Omega_{\rm gas}=0.5$, with
64 particles per pancake wavelength $\lambda_p$, to evolve the system 
to $a=2.333a_c$.
Figure 2 shows gas density, temperature, velocity, and pressure
profiles (in computational units) at the final time. 
The shock is located at $\log(x/\lambda_p)=-1.3$ (from midplane).
The ASPH results are significantly better than the SPH results,
which is
especially noticeable in the postshock density and temperature profiles.
The ASPH profiles follow the exact postshock solution over nearly 3 orders 
of magnitude in length, while the SPH profiles level off because of
limited spatial resolution.

\section{False Cooling}

A serious numerical problem 
emerges for both algorithms when radiative cooling is added.
In Figure~3, we plot a typical cooling function
commonly used in cosmological simulations. Let us
assume that the rectangle in the top panel of Figure~3 represents
the physically relevant region. The cooling rate drops sharply at low and 
high temperatures. To approximate this behavior, we consider
a simplified cooling window function illustrated in the bottom panel
of Figure~3. The cooling rate is constant inside a fixed temperature
window (Region II), and zero outside that window (Regions I and III).

In principle, when a fluid element is shock-heated from Region~I to Region~III,
it should cross the shock too fast to cool radiatively as it
passes thru Region~II. The numerical shock has a finite thickness
which is unphysically large, however, so the shock transit time,
$t_{\rm shock}$, can artificially exceed the physical cooling
time in Region~II, $t_{\rm cool}$, causing spurious radiative cooling.
Since $t_{\rm shock}$ scales like the shock thickness
divided by the preshock velocity,
it depends on resolution,
as the number of particles across the shock tends to be constant.
If the cooling rate is so large or the resolution so poor that
$t_{\rm shock}>t_{\rm cool}$, particles will be 
unable to cross Region II, as cooling forces them back to 
Region~I, and the shock does not form. 
We call this the {\it False Cooling Problem}.
 
\begin{figure}
\plotone{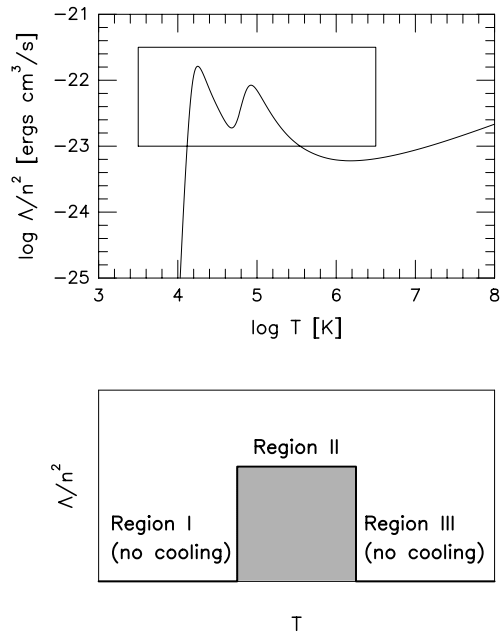}
\caption{Top panel: typical radiative cooling rate commonly used in
cosmological simulations. Bottom panel: top-hat cooling function
used to approximate the region of the cooling curve indicated by the
rectangular box in the top panel.}
\end{figure}

\begin{figure}
\plotone{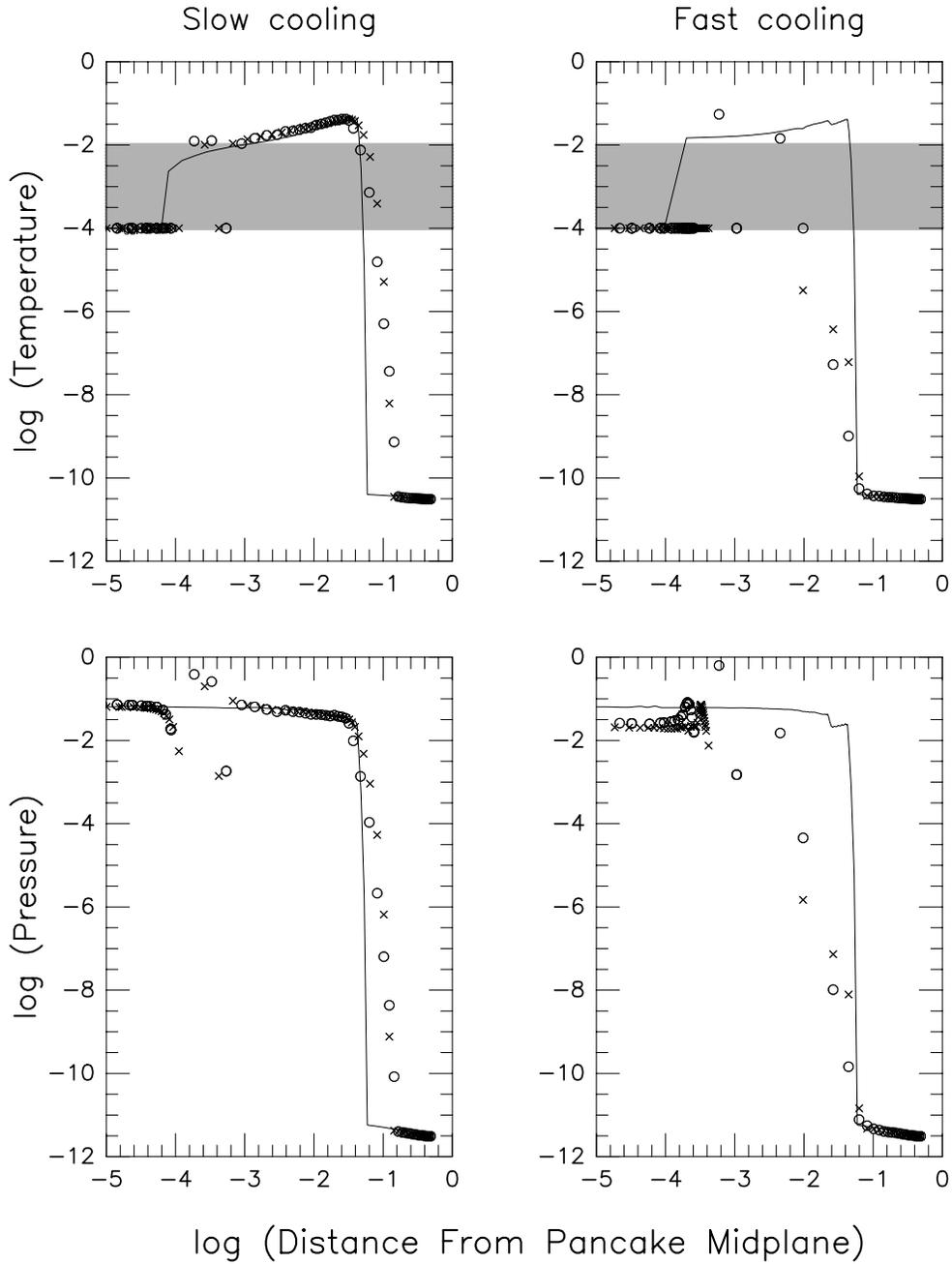}
\caption{Temperature and pressure
profiles at $a/a_c=2.333$ for pancake collapse with
radiative cooling. Shaded area indicates location of cooling
window. Left panels: slow cooling. Right panels: fast cooling.
Solid curves, crosses, and
circles show the correct solution
[by high-res., Lagrangian hydro method of Shapiro \& Struck-Marcell (1985)], 
the standard SPH results, and the ASPH 
results, respectively.
}
\end{figure}

\begin{figure}
\plotone{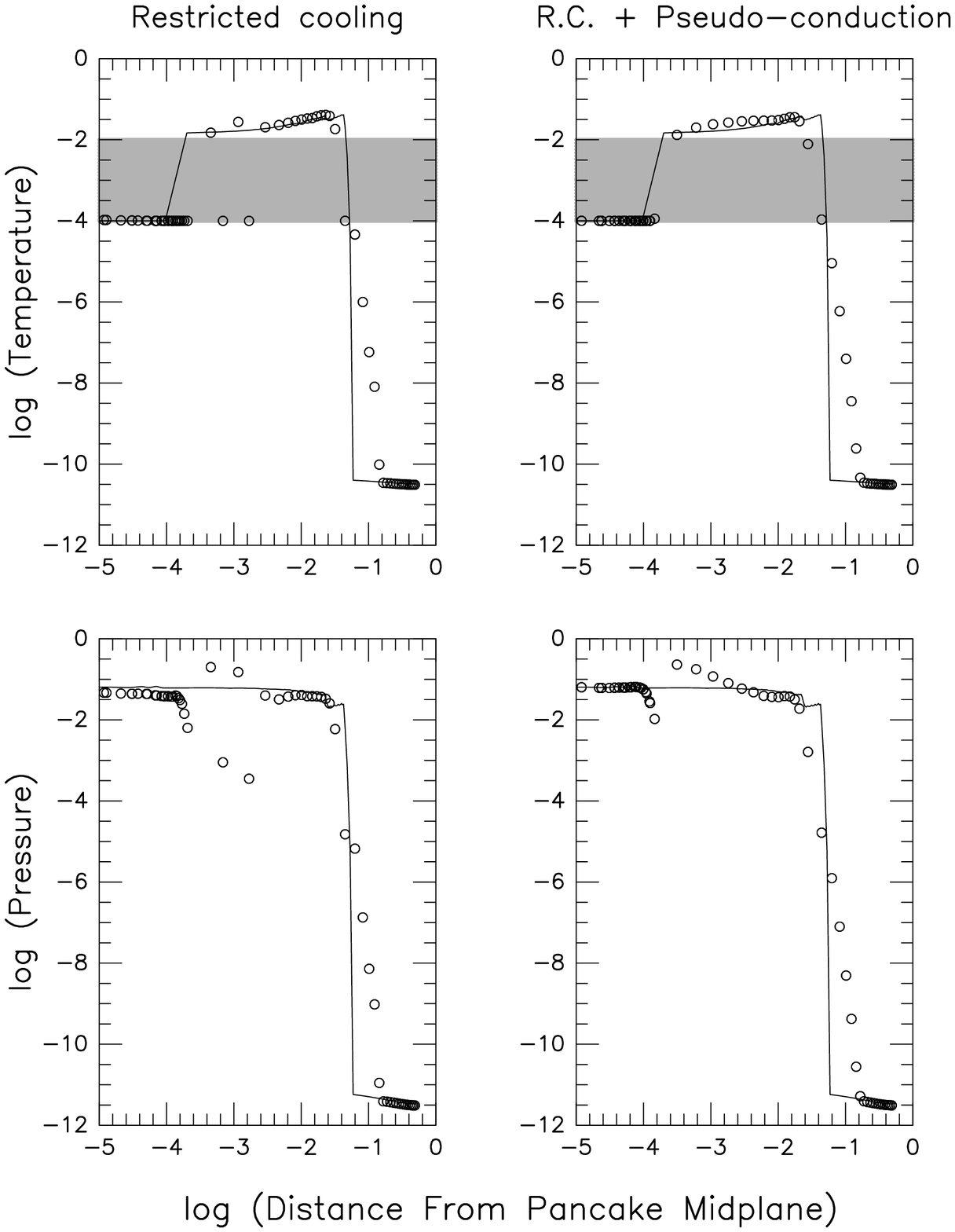}
\caption{Temperature and pressure profiles 
at $a/a_c=2.333$ for pancake collapse with
fast cooling, using ASPH. Shaded area indicates the location of the cooling
window. 
Left panels: no cooling while a particle shock-heats.
Right panels: no cooling while a particle shock-heats, 
and pseudo-conduction added.
}
\end{figure}

To illustrate this problem, we introduce a window cooling function to the 
cosmological pancake problem. Figure~4 shows the temperature and pressure
profiles at $a=2.333a_c$.
The shaded area indicates the cooling window. If cooling is slow
(left panels), the shocked
gas particles are able to cross the cooling window and
accumulate correctly
in the postshock region. The particles located at the bottom of
the cooling window are those whose postshock temperature was
inside the cooling window. They cooled {\it after} having been shocked.
When cooling is fast (right panels), however,
particles are incapable of reaching their correct postshock 
temperature, and the
shock does not even form.

\section{Possible Solution}

The false cooling problem
only affects 
particles that are going through a shock transition. If the algorithm
could track the location of shocks, then false cooling
could be eliminated, simply by not allowing particles undergoing
shock transitions to cool. There is already a shock-tracking
algorithm in ASPH, which is used to restrict viscous heating. 
The algorithm identifies particles located in shock transitions
and turns viscous heating on for them. We can attempt to solve
the false cooling problem by turning cooling off for the same particles,
with results as shown in the left panels of Figure~5. Particles are now
able to reach the postshock temperature and pressure 
(compare with the right panels of
Fig.~4), and the ASPH solution is close to the exact one. There are
some oscillations near the contact discontinuity between hot and
cooled gas, but since it involves only a few particles, these
oscillations are underresolved and do not feed back into the evolution of 
the system. The right panel shows an attempt to improve the solution by
introducing a small amount of artificial conduction. The oscillations are 
gone, but the postshock profile is not as well reproduced.

\section{Summary}

Anisotropic smoothing kernels
enable the ASPH algorithm to simulate problems involving
strongly anisotropic collapse or expansion more
accurately than isotropic SPH, with better length resolution
for the same number of particles.
Additional improvement results because ASPH tracks the location of shocks
and restricts viscous heating to particles overtaken by shocks.

We have identified a major problem, called {\it false cooling}, which can
affect cosmological simulations whenever
the cooling rate has temperature peaks. False cooling
prevents shocked gas from reaching regions of high temperature where the
cooling is low. Solving this problem may require a shock-tracking
capability like that of ASPH.
Our preliminary results are encouraging.

\acknowledgments

We thank Ilian Iliev for help with our Lagrangian hydro simulations.
This work was supported by TARP Grant 3658-0624-1999 
and NASA Grants NAG5-10825 and NAG5-10826.

\end{document}